\documentclass[onecolumn, preprint, aps, prb]{revtex4}
\usepackage{mathrsfs}
\usepackage[dvips]{graphicx}
\usepackage{latexsym}
\usepackage{amsmath}
\usepackage{subfigure}
\usepackage{amsfonts}
\begin{document}

\title{Scaled Particle Theory for Hard Sphere Pairs. II. Numerical  Analysis}
\author{Swaroop Chatterjee}
\affiliation{Department of Chemical Engineering, Princeton University, Princeton, NJ 08544, USA}
\author{Pablo G. Debenedetti}
\affiliation{Department of Chemical Engineering, Princeton University, Princeton, NJ 08544, USA}
\author{Frank H. Stillinger}
\affiliation{Department of Chemistry, Princeton University, Princeton, NJ 08544, USA}
\date{\today}
\begin{abstract}
We use the extension of scaled particle theory (ESPT) presented in
the accompanying paper [Stillinger et al. J. Chem. Phys. xxx, xxx
(2007)] to calculate numerically pair correlation function of the
hard sphere fluid over the density range $0\leq \rho\sigma^3\leq
0.96$. Comparison with computer simulation results reveals that the
new theory is able to capture accurately the fluid's structure
across the entire density range examined. The pressure predicted via
the virial route is systematically lower than simulation results,
while that obtained using the compressibility route is lower than
simulation predictions for $\rho\sigma^3\leq 0.67$ and higher than
simulation predictions for $\rho\sigma^3\geq 0.67$. Numerical
predictions are also presented for the surface tension and Tolman
length of the hard sphere fluid.
\end{abstract}
\maketitle

\section{Introduction} During roughly the last half century, the statistical mechanics of many-body systems has
incorporated and has benefited from various forms of computer
simulation.  Indeed, with the remarkable advances in computer power
that have occurred during that period, simulation has grown in the
depth and breadth of its contributions to a point of near dominance
over more analytical approaches.  The hard sphere model provides a
clear example, where its rich repertoire of both equilibrium and
non-equilibrium phenomena are now better appreciated and understood
as a result of a wide range of simulation activities~\cite{Rosen1,
Alder2, Alder1, Hoover1, Rintoul1}.

Nevertheless, continuing to pursue the more analytical aspects of
many-body statistical mechanics remains a valuable complementary
mission. This and the immediately preceding companion
paper~\cite{Stillinger1} (hereafter referred to as ``I"), fall
into this more analytical category. The objective has been to deepen
understanding of the equilibrium properties of the classical hard
sphere system. The underlying strategy involves an extension (ESPT)
of the "scaled particle theory" (SPT) originally presented by Reiss,
Frisch, and Lebowitz~\cite{Reiss1}. The mathematical strategy of
our approach has been described in "I". The present paper ``II"
provides details of a numerical investigation based on the
mathematical formalism developed in ``I". One of our principal
objectives is numerical prediction of the hard sphere pair
correlation function $g(r,\rho)$ over a substantial range in density
$\rho$.

The following Section II describes the coupled equations that must be self-consistently solved. At the center of this
procedure is a non-linear integral equation for $g(r,\rho)$. Successful solution of this equation at high density
demands use of somewhat sophisticated techniques that are explained in Section II. Section III presents results not only
for the basic quantity $g(r,\rho)$, but also for the pressure and surface free-energy properties of the hard sphere
system. This investigation has inevitably raised some important questions, particularly about what might be the most
profitable directions for future extensions of the approach.  These questions and our conclusions based on the results
obtained thus far form the subjects for Section IV.

\section{Numerical procedure} The goal of our numerical calculations is to obtain the hard sphere pair correlation
function $g(r,\rho)$. In ``I", we introduced the overall general
strategy behind this calculation. Here, we outline the solution
procedure in greater detail. The computation can be divided into two
major parts. The first involves the calculation of the isothermal
reversible work $W(\lambda,\rho)$ required to create a single
spherical cavity in a hard sphere fluid, where $\lambda$ is the
cavity radius, and the second involves the calculation of the double
or bi-spherical isothermal and reversible cavity formation work
$W_2(r,\lambda,\rho)$, where $r$ is the separation between the two
cavity centers. By calculating the work done in these two cases, the
pair correlation function can be obtained using the following
relation ($\beta=1/k_BT$):
\begin{equation}
\label{gr}
 g(r,\rho)=\exp[-\beta W_2(r,1,\rho)+2\beta W(1,\rho)]
\end{equation}

To use the above relation, we note that the cavity work terms
$W(\lambda,\rho)$ and $W_2(r,\lambda,\rho)$ must be evaluated at
$\lambda=\sigma$, the hard sphere diameter (i.e., $\lambda=1$, using
$\sigma$ as the natural scale for distance). Using the extended
scaled particle theory (ESPT) of ``I", the single cavity work is
described by the following set of equations for different ranges of
$\lambda$:
\begin{eqnarray}
\beta W(\lambda,\rho) = &-\ln\bigg(1-4\pi\rho\lambda^3/3\bigg)&
 ,\bigg(0\leq\lambda\leq\frac{1}{2}\bigg) \\
 =& -\ln\bigg(1-4\pi\rho\lambda^3/3 +
 2\pi^2\rho^2\int_1^{2\lambda}dr\bigg[4\lambda^3r^2/3-\lambda^2r^3\nonumber \\
 & +r^5/12\bigg]g(r,\rho)\bigg) &,\bigg(\frac{1}{2}\leq\lambda\leq3^{-1/2}\bigg)  \label{Wlt3}\\
 =&
 J(\rho)\lambda^3+K(\rho)\lambda^2+L(\rho)\lambda+M(\rho)+N(\rho)/\lambda &
,\bigg(\frac{1}{3^{1/2}}\leq\lambda\bigg) \label{Wgt3}
\end{eqnarray}
Therefore, the calculation of $W(\lambda=1,\rho)$ requires finding
the coefficients $J\ldots N$ such that $W(\lambda,\rho)$ is a smooth
function of $\lambda$. Among these, the coefficient $J$ is related
to the pressure $P$, $K$ to the surface tension $\gamma$ and $L$ to
its curvature dependence. The fitting of these coefficients is
performed at $\lambda=3^{-1/2}$. In other words, we use the
expressions given in the equations \ref{Wlt3} and \ref{Wgt3} to
calculate values of the coefficients that ensure a smooth transition
across $\lambda=3^{-1/2}$. This is accomplished by equating the work
function value, along with its first and second
$\lambda-$derivatives, in the two regions (equations \ref{Wlt3} and
\ref{Wgt3} ) providing constraints on three of the five
coefficients. The other two necessary conditions are obtained by
evaluating the pressure $P$, and hence $J(\rho)$ from the pair
correlation function
\begin{equation}
J(\rho) = \frac{4\pi}{3}\beta
P=\frac{4\pi\rho}{3}+\frac{8\pi^{2}\rho^{2}}{9}g(1,\rho)
\end{equation}
and the following condition arising from the density of particle
centers $\rho g(r,\rho)$ at the surface of the $\lambda=1$ cavity
\begin{equation}
\beta W'(r,\lambda=1) = 4\pi\rho g(1,\rho)
\end{equation}
We note here that in this $W(\lambda,\rho)$ calculation, $g(r,\rho)$
is required as input to evaluate the necessary conditions. We will
further address this issue later on in this section.

Having obtained $W(\lambda=1,\rho)$, we turn our attention to the
bi-spherical cavity work $W_2(r,\lambda,\rho)$. This calculation
follows along similar lines as in the single cavity case. In ``I",
we developed expressions for the $W_2(r,\lambda,\rho)$ for two
regimes: one where the range of cavity sizes ($\lambda$) limits the
maximum possible number of clusters of spheres of unit diameter that
can be found inside the bi-spherical cavity to 2, and the other
where the maximum occupancy is greater than 2. We display them here
in a different form.
\begin{eqnarray}
\beta W_2(r,\lambda,\rho)&=& -\ln\bigg\{1-\rho V(r,\lambda) +\frac{2\pi^2\rho^2}{r}\int_{r-2\lambda}^{r}dr_1\nonumber\\
& &\bigg(\frac{8}{15}\lambda^5-\frac{2}{3}\lambda^3(r-r_1)^2+\frac{1}{3}\lambda^2(r-r_1)^3-\frac{1}{60}(r-r_1)^5\bigg)
r_1 g(r_1) + \frac{2\pi^2\rho^2}{r}\int_{r}^{r+2\lambda}dr_2 \nonumber\\
& &\bigg(\frac{8}{15}\lambda^5-\frac{2}{3}\lambda^3(r-r_2)^2-\frac{1}{3}\lambda^2(r-r_2)^3+\frac{1}{60}(r-r_2)^5\bigg)
r_2 g(r_2)\bigg\} ,\:for\:n=2 \label{W2eq2}\\
\beta
W_2(r,\lambda,\rho)&=&\frac{3J(\rho)}{4\pi}V(r,\lambda)+\frac{K(\rho)}{4\pi}A(r,\lambda)+X(\rho,s)\lambda+Y(\rho,s)+\frac{Z(%
\rho,s)}{\lambda},\:for\:n\geq2 \label{W2gt2}
\end{eqnarray}
where $V(r,\lambda)$ and $A(r,\lambda)$ are the volume and surface
area of the double cavity, and $s$ is a transformation variable
defined as $s=r-2\lambda$. Physically, $s$ describes the distance
between the nearest points on the surface of the two spherical
cavities, when they are disconnected, and the corresponding distance
between the interpenetrating surfaces, when they overlap. In the
former case, $s$ is positive and it is negative in the latter case.
Using this variable allows us to circumvent singularities arising
due to the disconnection of the two cavities of size $\lambda$ at
$r=2\lambda$.

The boundary between the two regimes, as discussed in ``I", for
$r\geq (3^{1/2}-1)/2$ is at $\lambda=1/2$, and for $0\leq r<
(3^{1/2}-1)/2$ is given by
$\lambda=(1/2)(3^{1/2}-r)+(1/8)(3^{1/2}-r)^{-1}$, where we have
expressed distances in units of $\sigma$. We assume that equation
\ref{W2gt2} holds for all $r$, provided $\lambda>1/2$. Furthermore,
coefficients $J(\rho)$ and $K(\rho)$ are the same as those obtained
in the single cavity $W(\lambda,\rho)$ calculation. Therefore, there
are three coefficients that need to be determined, namely
$X(\rho,s)$, $Y(\rho,s)$ and $Z(\rho,s)$. These are determined by
equating the value of the work function $W_2$, and its first and
second derivatives with respect to $\lambda$ at $\lambda=1/2$.

Along with the transformation variable $s$, another variable
$t=2\lambda-1$, a rescaled and shifted cavity size, is also used. In
these variables, the fitting described above is done for each $s$
value at $t=0$. The pair correlation calculation in equation
\ref{gr} therefore requires that the coefficients $X$, $Y$ and $Z$
be determined for various values of the intercavity distance $s$ at
$t=1$. The range of $s$ in this calculation is $-1\leq s <\infty$,
which corresponds to $1\leq r <\infty$.

Once all eight coefficients $J(\rho)$, $K(\rho)$, $L(\rho)$,
$M(\rho)$, $N(\rho)$, $X(\rho,s)$, $Y(\rho,s)$, and $Z(\rho,s)$ are
computed, the calculation of $g(r,\rho)$ follows from equation
\ref{gr} by setting $\lambda=1$ in the single cavity work
$W(\lambda,\rho)$ calculation and $t=1$ in the double cavity
$W_{2}(s,t,\rho)$ calculation. However, as noted earlier in this
section, both work calculations require $g(r,\rho)$ as input(see
equations \ref{Wlt3} and \ref{W2eq2}). To accomplish this pair
correlation function calculation, a numerical solution technique is
employed by using an estimate of $g(r,\rho)$ as input, and then
iteratively correcting it after calculating $g(r,\rho)$ using the
procedure discussed above. An arbitrarily small tolerance is chosen
to determine whether the calculation has converged by comparing the
the calculated and input $g(r,\rho)$ functions.

An appropriate numerical method to correct the input $g(r,\rho)$
using the calculated values was determined after some trials on our
part. The first attempted technique simply used a linear combination
of the $n^{th}$ step input and calculated values in some arbitrary
proportion to obtain the $(n+1)^{th}$ step input $g(r,\rho)$. In
other words,
\begin{equation}
g^{input}_{n+1}(r,\rho)=x\times g^{input}_n(r,\rho)+(1-x)\times
g^{calculated}_n(r,\rho)
\end{equation}
where $x$ defines the proportion of calculated and input value in
the corrected input $g(r,\rho)$ for the next stage. This method is
only effective at low densities ($\rho<0.4$; here and in the
remainder of the paper, dimensionless densities are obtained by
scaling with $\sigma$, the hard sphere diameter), regardless of the
value assigned to $x$. The reason for this might be the high degree
of non-linearity in the problem, which makes it difficult to achieve
convergence using such a crude technique if the initial input
$g(r,\rho)$ is not an accurate enough estimate of the actual
function. In this connection, we used, as two alternative input
$g(r,\rho)$'s, a polynomial extrapolation of the converged pair
correlation functions obtained for densities lower than the one
being computed, or the converged pair correlation function of a
lower density case. These changes are able to improve the overall
speed and convergence of the calculation. However, beyond a density
$\rho\cong0.6$, even these methods are unable to ensure convergence
of the calculation.

Satisfactory behavior in both convergence and speed was eventually
obtained by using a generalized minimum residual method (GMRES) to
solve the system of equations. The conventional GMRES method was
developed for systems of linear equations ~\cite{Saad1}. It can be
shown that for linear equations, the solution can be approximated by
the $n^{th}$ iterate of the vector in a Krylov subspace (defined
below) which minimizes the residual. Mathematically, for a system of
equations of the form $Ax=b$, where $x$ is the variable to be
solved, the $n^{th}$ Krylov subspace is given by
\begin{equation}
K_{n}=span\{b,Ab,A^{2}b,\ldots,A^{n-1}b\}
\end{equation}
where $span\{z_1,z_2,\ldots\}$ refers to the set of linearly
dependent vectors which can be constructed from a vector set $z_1$,
$z_2,\ldots$ The solution using the GMRES method is a vector in this
subspace $x_{n}\epsilon K_{n}$ such that the norm of residual
$Ax_{n}-b$ is minimized as part of a least-squares problem. In order
to solve this problem, we need to find the orthonormal basis vectors
of $K_{n}$. The component vectors in the above equation ($b$, $Ab$,
$\ldots$, $A^{n-1}b$), being linearly dependent, are therefore not
suitable to form this basis vector set. Instead, the Arnoldi
process~\cite{Kelley1} is used to construct the necessary
orthonormal basis vectors $v_{1}$, $\ldots$, $v_{n}$ of $K_{n}$.
This process also produces an upper Hessenberg matrix $H_{n}$ of
dimensions $n+1$ by $n$ such that
\begin{equation}
Av_{n}=v_{n+1}H_{n}
\end{equation}
The orthogonality of the basis vectors leads to a simplification in the least-squares problem, or in other words the
objective function is altered as
\begin{equation}
\min_{x_{n}\epsilon K_{n}}
\parallel Ax_{n}-b
\parallel \equiv \min_{y_{n}\epsilon \mathbb{R}^{n}}
\parallel H_{n}y_{n}-e_{1}
\parallel
\end{equation}
where
\begin{equation}
\label{xy} y_{n}=x_{n}/v_{n}
\end{equation}
and $e_{1}=(1,0,0,\ldots,0)$ is the first standard basis vector of
$\mathbb{R}^{n}$. At each step in the solution procedure, GMRES
solves for $y_{n}$ and then uses equation \ref{xy} to find $x_{n}$.
This process is repeated with increasing value of $n$ until the
residual $\parallel H_{n}y_{n}-e_{1}\parallel$ is within an assigned
tolerance value.

This general approach, although originally developed for systems of
linear equations, has  been extended to handle nonlinear
equations~\cite{Bell1}. The main difference lies in the
objective function. In solving a system of nonlinear equations
$F(x)=0$, the least squares problem at each step in the calculation
is formulated as  $\min_{s_{n}\epsilon K_{n}}\parallel
F(x_{n})+F'(x_{n})s_{n}\parallel$ where $s_{n}$ is related to
$x_{n}$ by the equation
\begin{equation}
s_{n}=s_{0}+x_{n}
\end{equation}
where $s_{0}$ is a chosen parameter. The GMRES method can be shown
to find a solution to a linear system of equations in at most $n$
iterations~\cite{Saad1}. For the nonlinear system of equations,
however, no such guarantees can be made. As a result, a maximum
number of iterations ($n_{max}$) is set. If the residual is not
within the set tolerance value after $n_{max}$ iterations, the
calculation is restarted such that $s_{0}=s_{n_{max}}$. A maximum
number of restarts is also used as input to the method so that the
calculation can be terminated in the event that convergence is not
achieved. The maximum distance $r_{max}$ was set at $r=9$ for these
$g(r,\rho)$ calculations. The computation interval $1\leq r \leq
r_{max}$ was divided into 2000 discrete intervals, and Simpson's
rule was used in all integrations.

The major thrust of this paper has been to calculate the pair
correlation function $g(r,\rho)$ for a hard sphere fluid based on
the extended scaled particle theory (ESPT) of ``I". Using the
methods discussed in this and the previous section, $g(r,\rho)$ has
been computed for different densities. As an initial estimate for
each $g(r,\rho)$ calculation, the converged $g(r,\rho)$ for a lower
density case was used. A tolerance limit of $10^{-7}$ was set for
each calculation to determine whether the residual was low enough.

\section{Results} 
\begin{figure}[!h]
     \centering
     \subfigure[]{
          \label{0.2}
          \includegraphics[width=8.8cm]{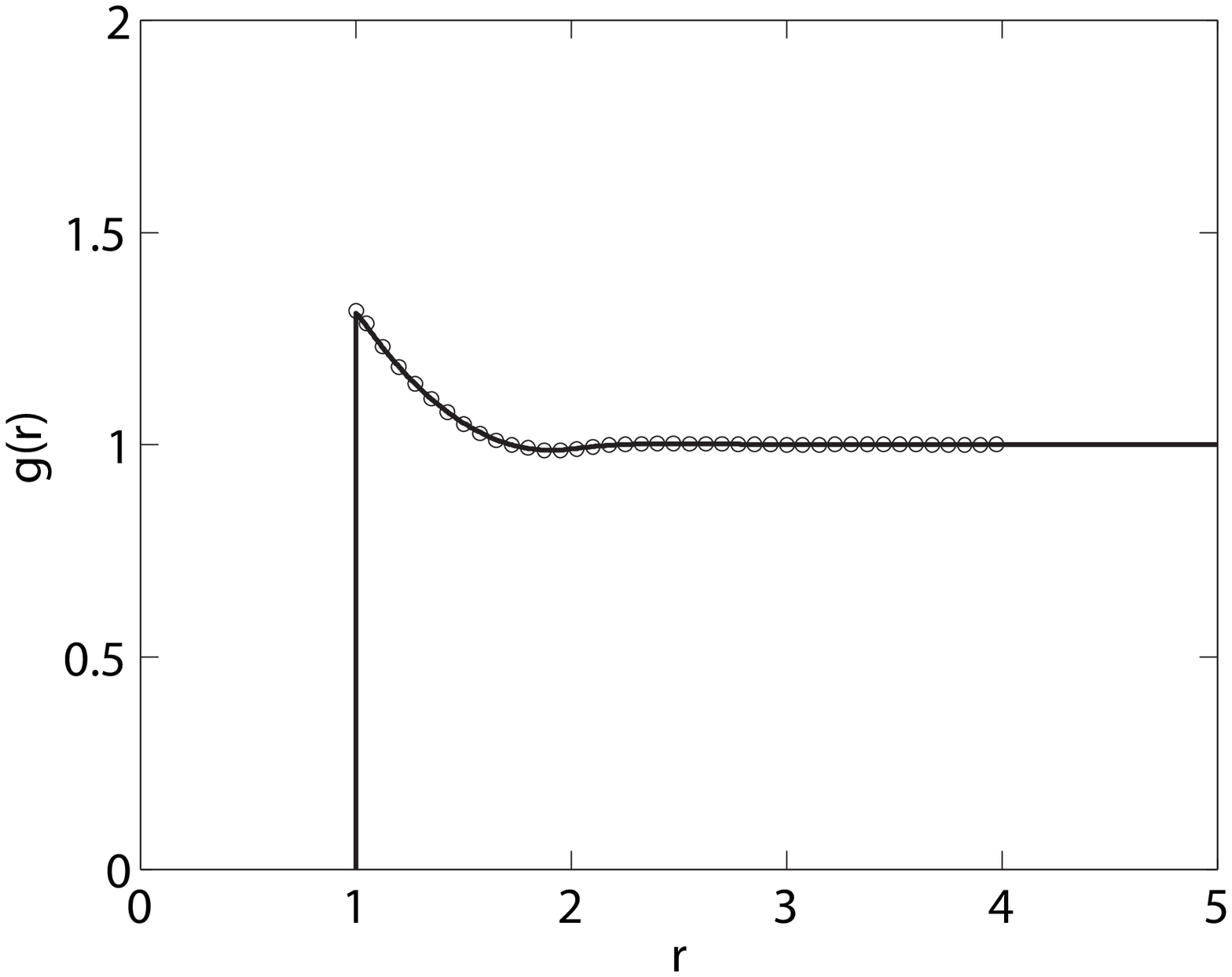}}
          \hspace{.3in}
     \subfigure[]{
          \label{0.3}
          \includegraphics[width=7.8cm]{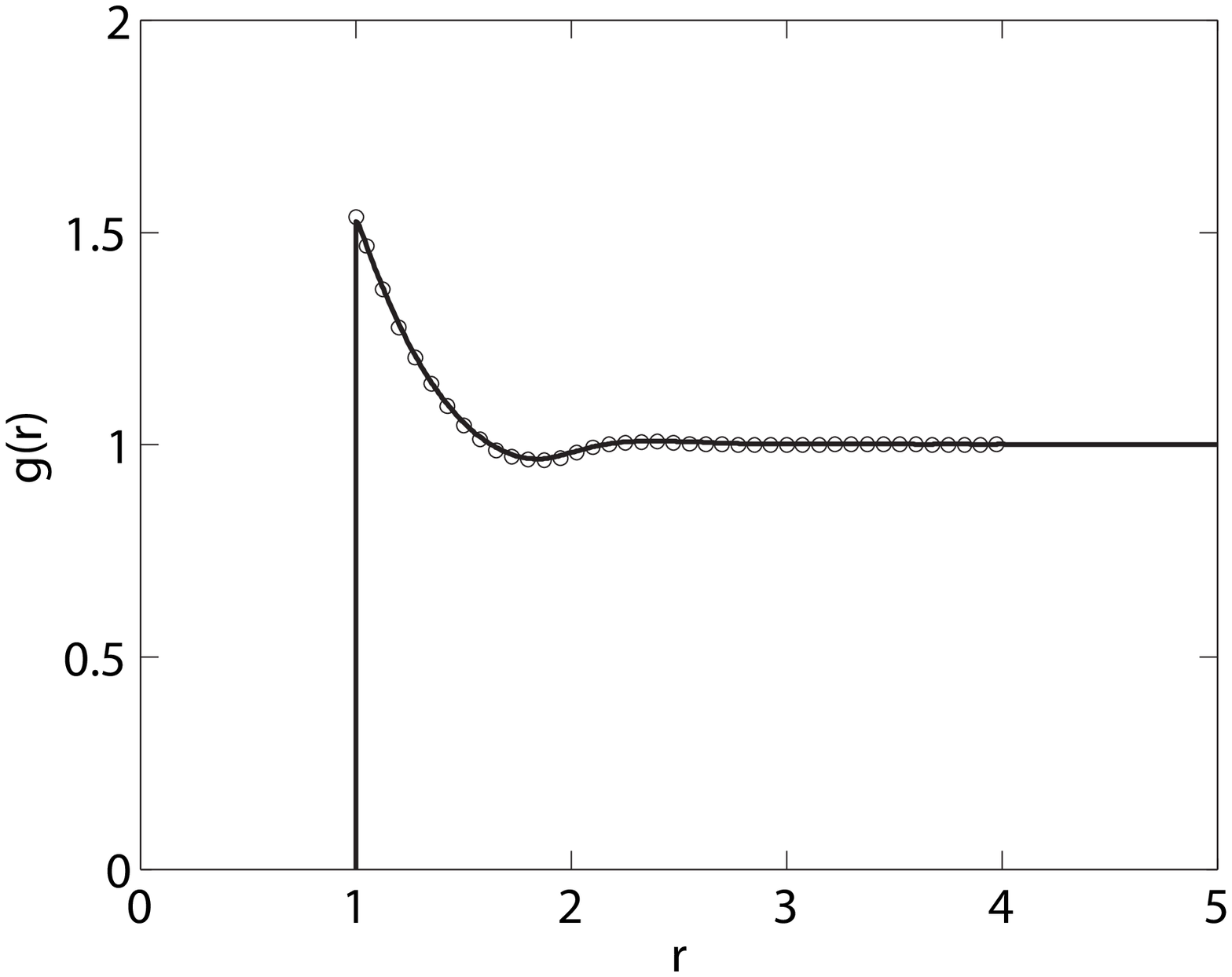}}
          \hspace{.1in}
    \subfigure[]{
          \label{0.4}
          \includegraphics[width=7.8cm]{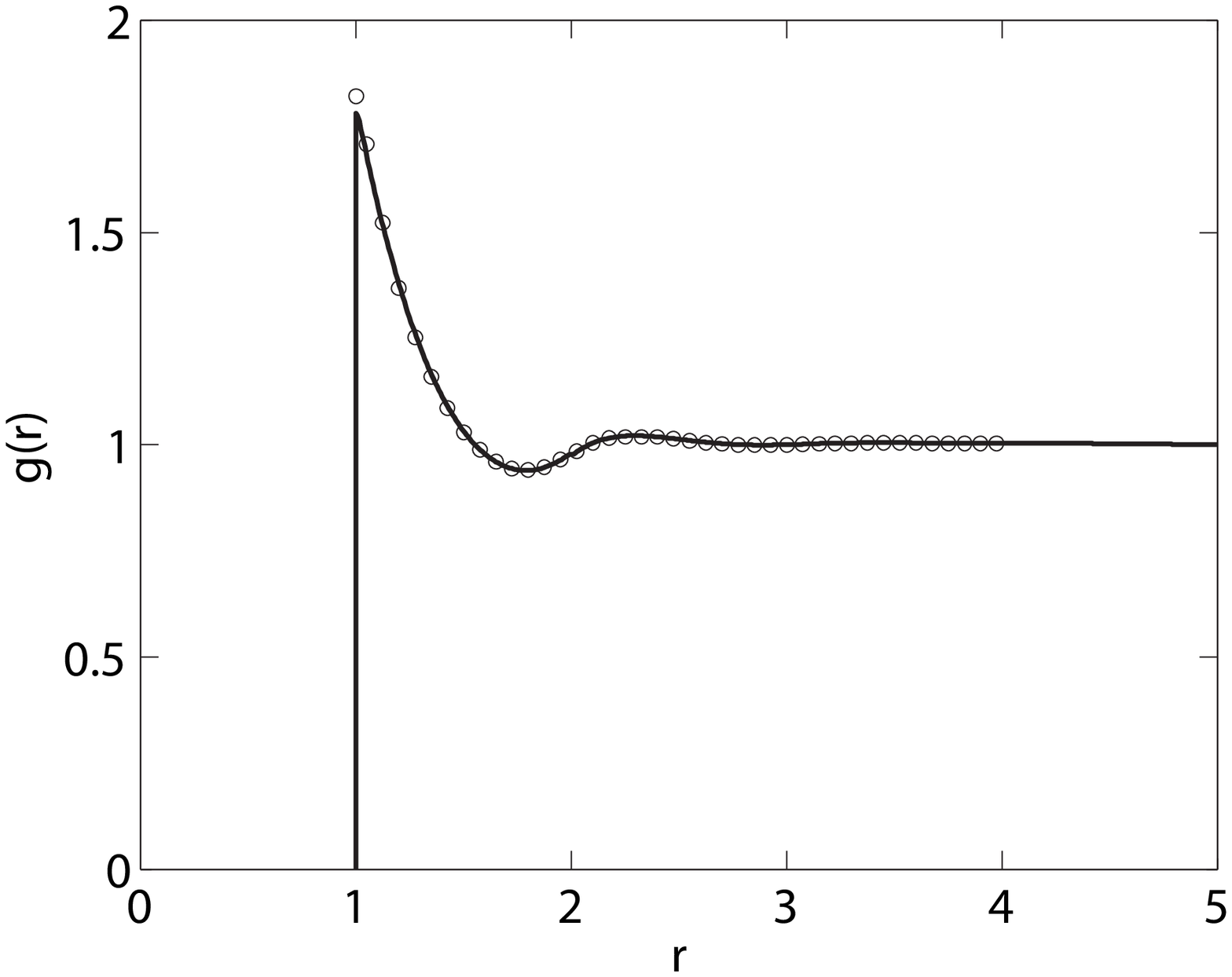}}
    \caption{The pair correlation function obtained using the extended scaled particle theory (ESPT). Here, three densities are shown: (a) $g(r,\rho=0.2)$, (b) $g(r,\rho=0.3)$ (c) $g(r,\rho=0.4)$. The lines are ESPT calculations. The open circles are molecular dynamics simulations~\cite{Lab1}. Distances have been scaled with the hard sphere diameter $\sigma$.}
    \label{g_r1}
\end{figure}

\begin{figure}[!h]
     \centering
     \subfigure[]{
          \label{0.5}
          \includegraphics[width=8.8cm]{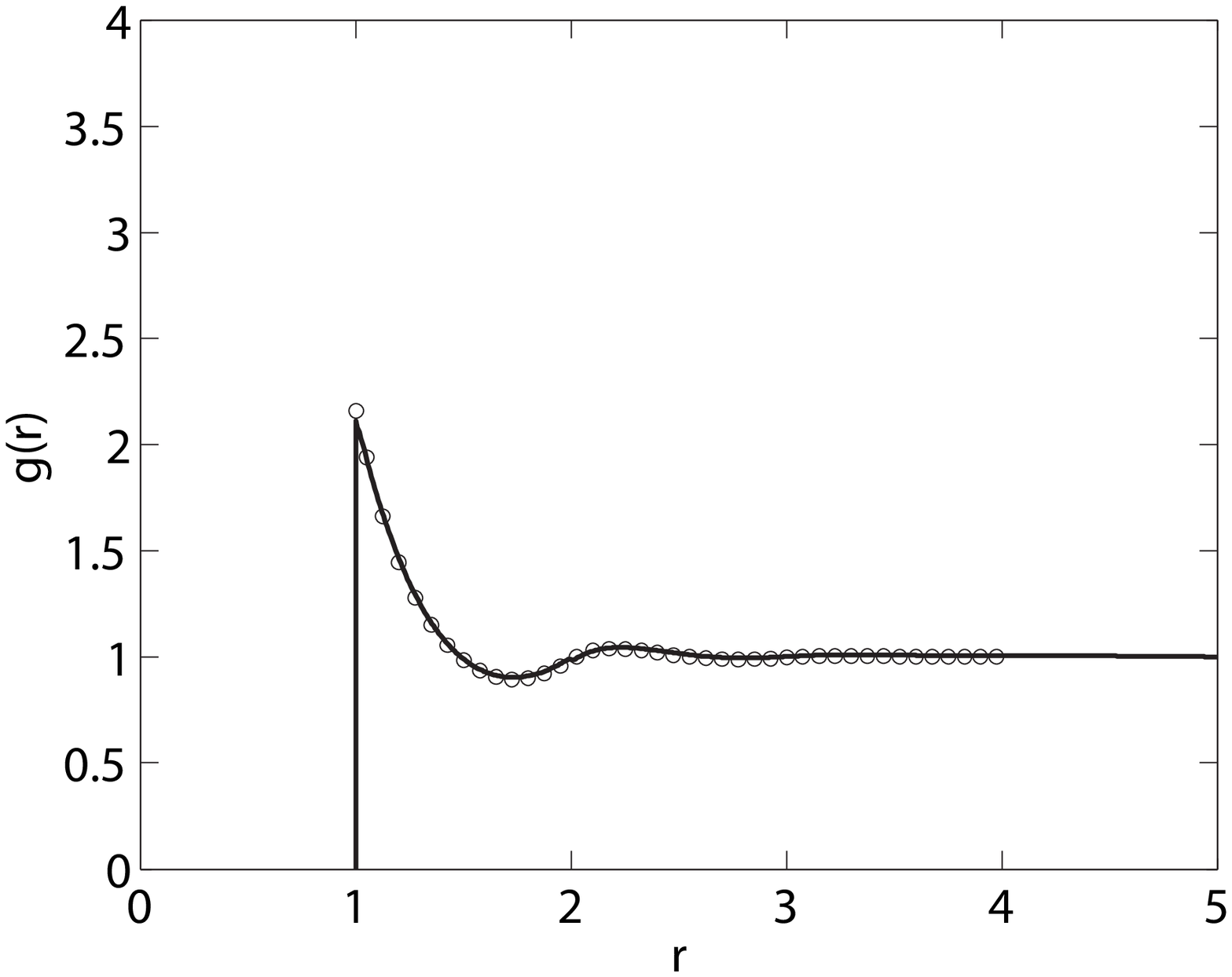}}
          \hspace{.3in}
     \subfigure[]{
          \label{0.6}
          \includegraphics[width=7.8cm]{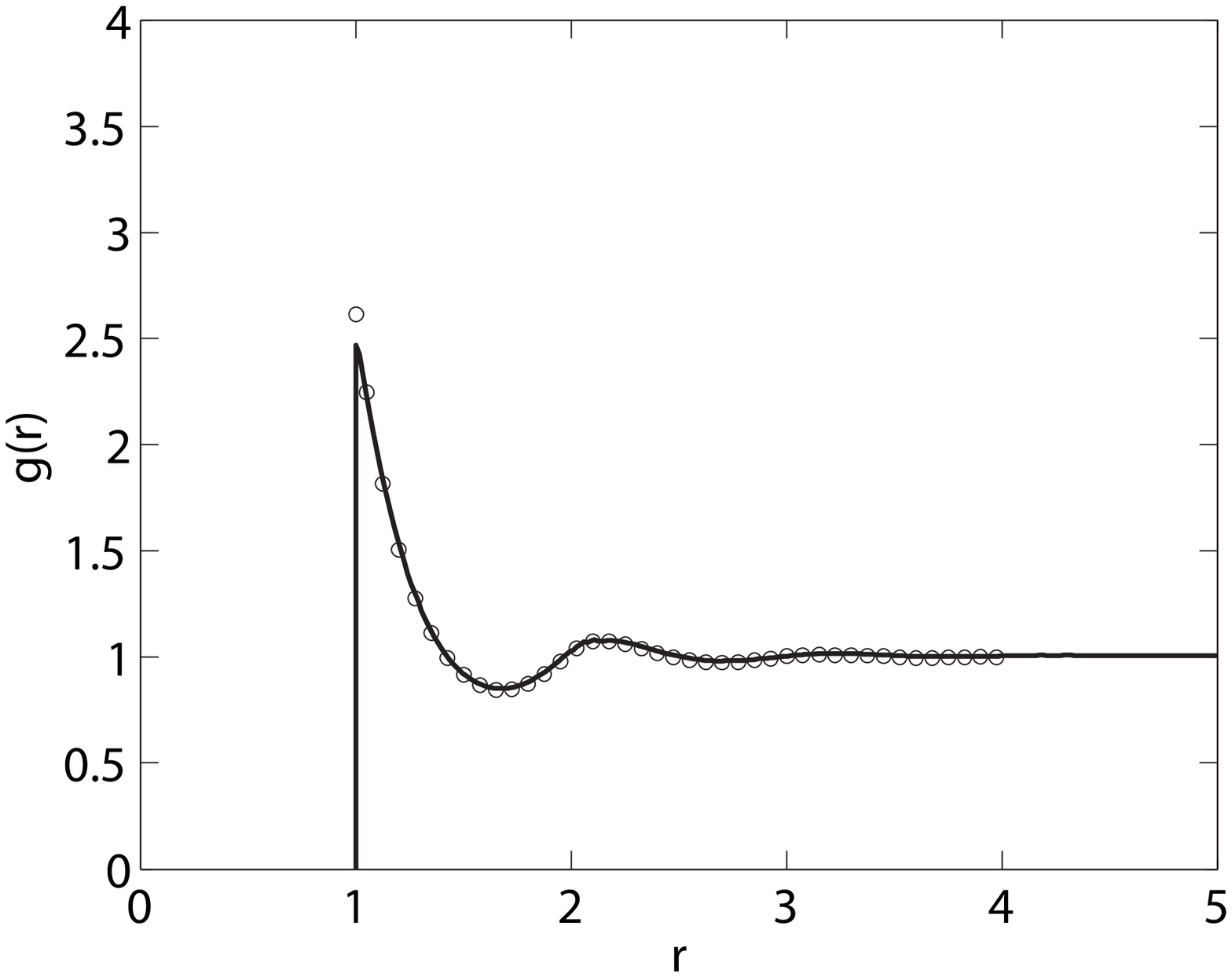}}
          \hspace{.1in}
    \subfigure[]{
          \label{0.7}
          \includegraphics[width=7.8cm]{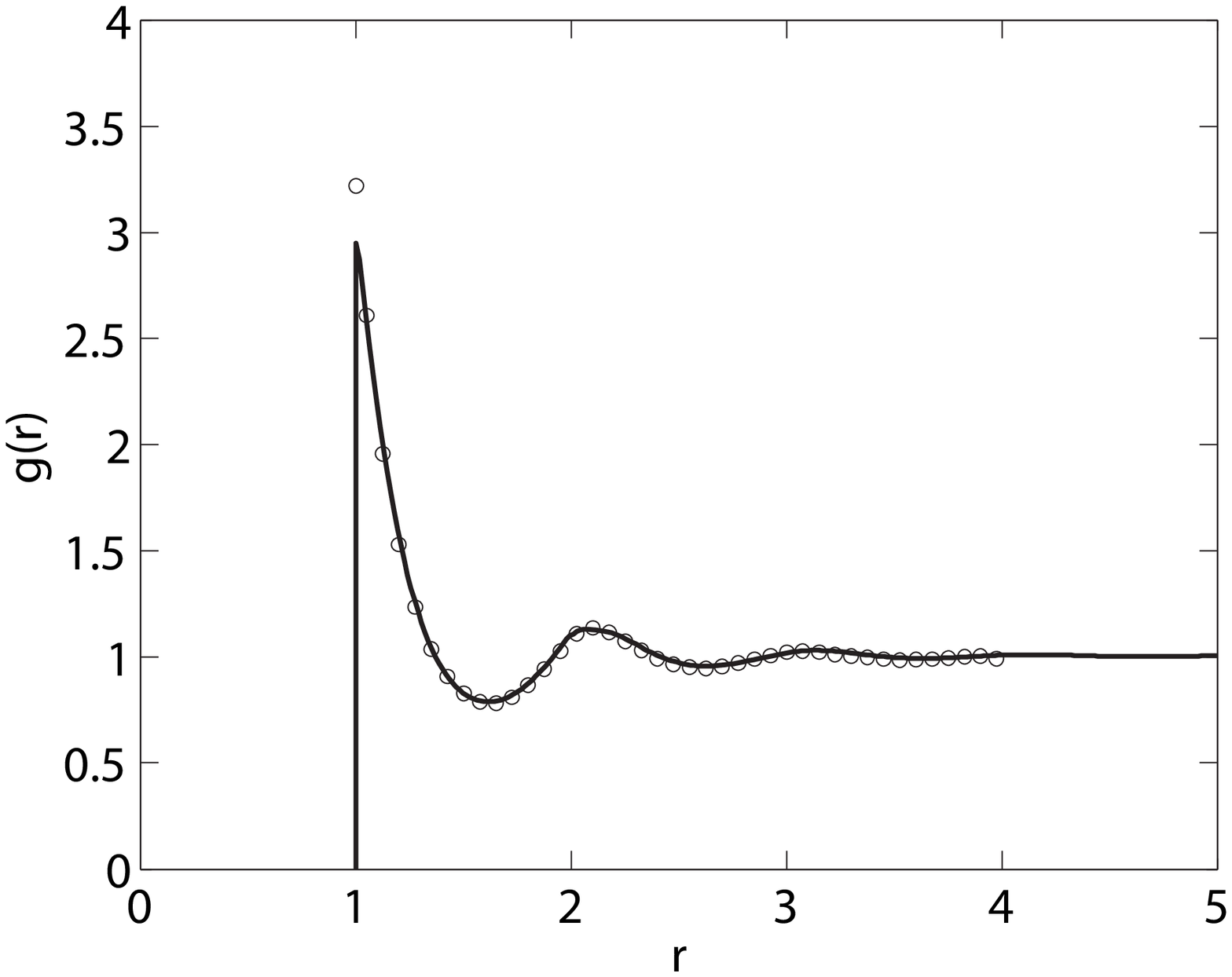}}
    \caption{The pair correlation function obtained using the extended scaled particle theory (ESPT). Here, three densities are shown: (a) $g(r,\rho=0.5)$, (b) $g(r,\rho=0.6)$ (c) $g(r,\rho=0.7)$. The lines are ESPT calculations. The open circles are molecular dynamics simulations~\cite{Lab1}. Distances have been scaled with the hard sphere diameter $\sigma$.}
    \label{g_r2}
\end{figure}

\begin{figure}[!h]
     \centering
     \subfigure[]{
          \label{0.8}
          \includegraphics[width=8.8cm]{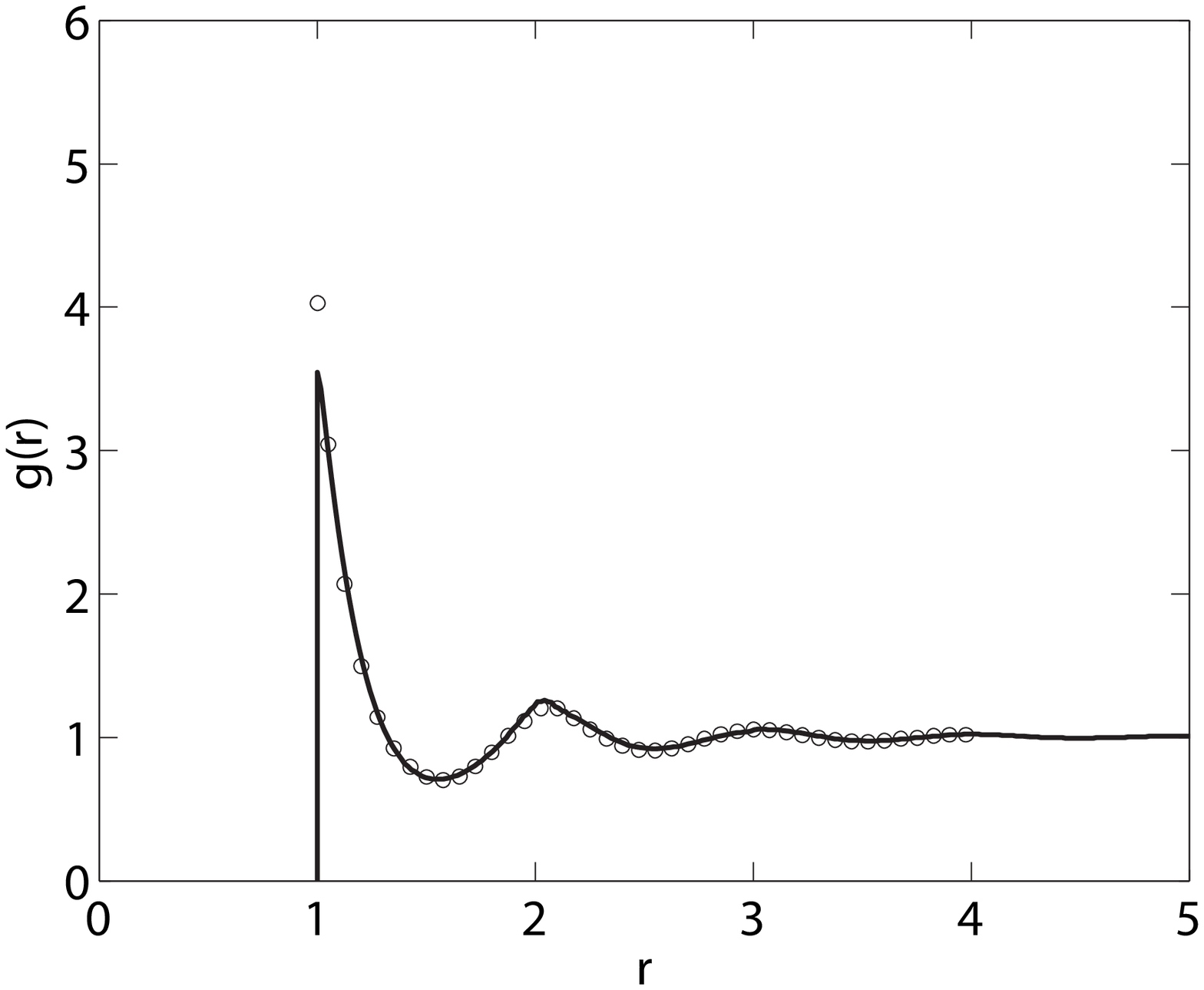}}
          \hspace{.3in}
     \subfigure[]{
          \label{0.9}
          \includegraphics[width=7.8cm]{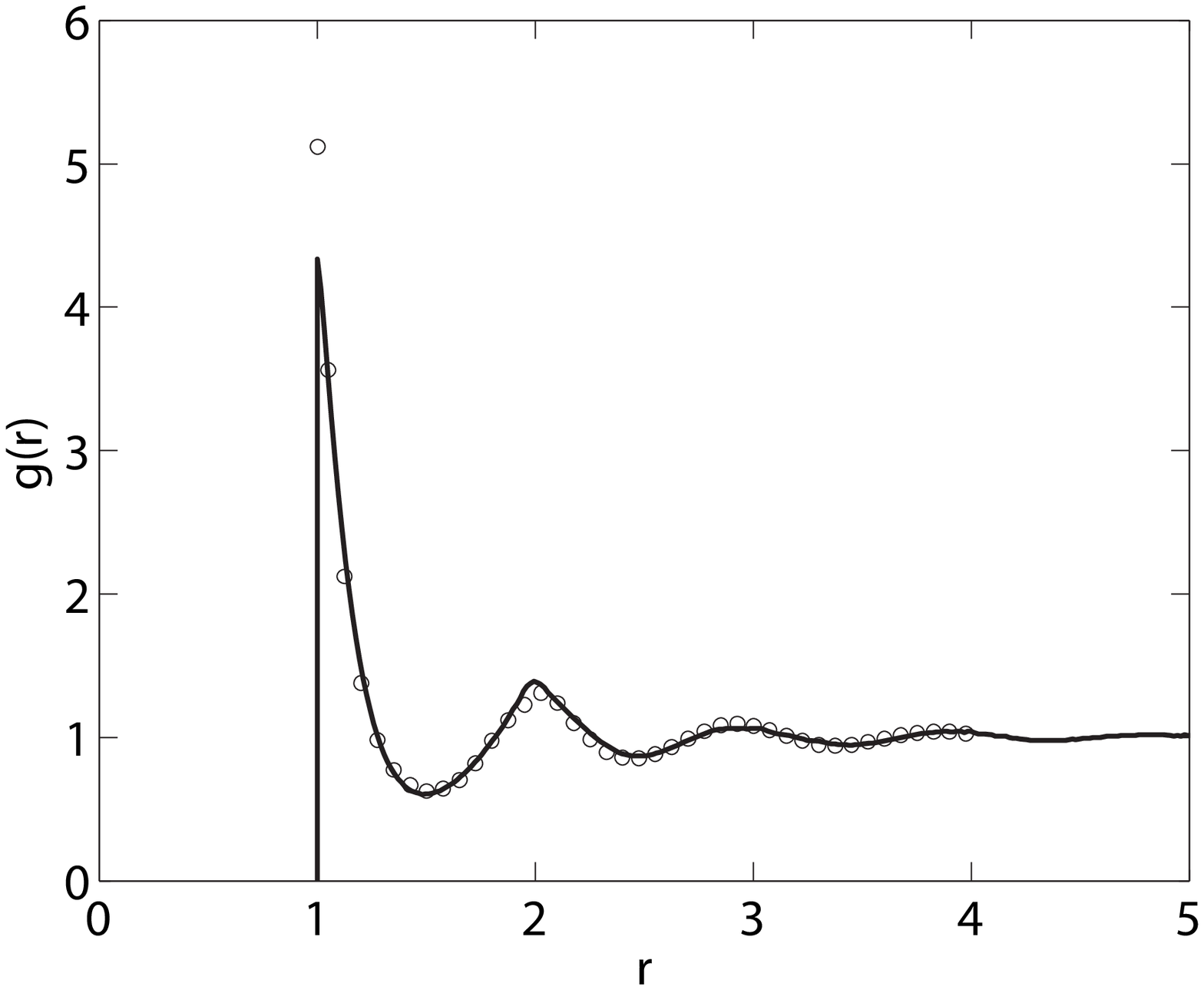}}
          \hspace{.1in}
    \subfigure[]{
          \label{0.96}
          \includegraphics[width=7.8cm]{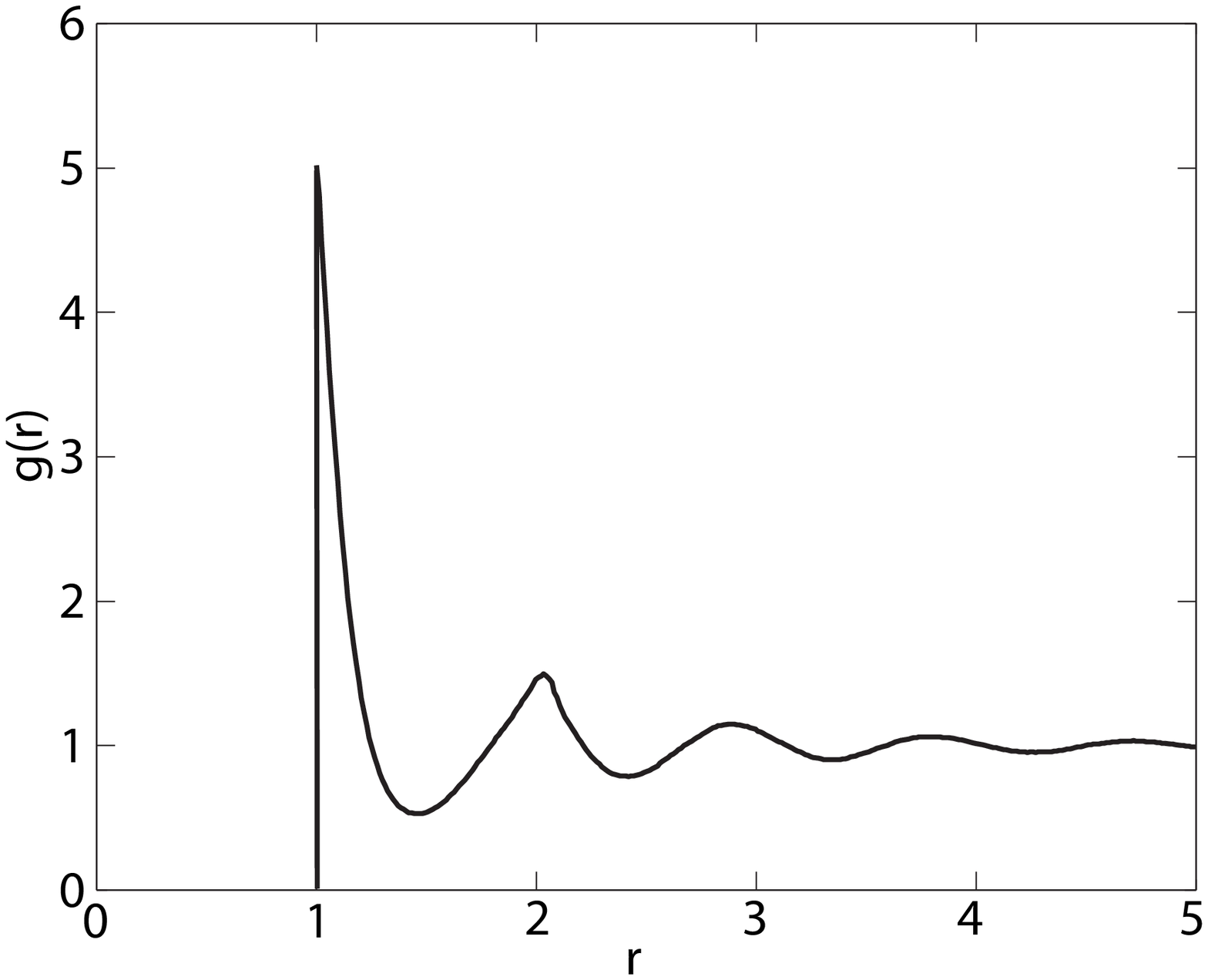}}
    \caption{The pair correlation function obtained using the extended scaled particle theory (ESPT). Here, three densities are shown: (a) $g(r,\rho=0.8)$, (b) $g(r,\rho=0.9)$ (c) $g(r,\rho=0.96)$. The lines are ESPT calculations. The open circles are molecular dynamics simulations~\cite{Lab1}. Distances have been scaled with the hard sphere diameter $\sigma$.}
    \label{g_r3}
\end{figure}

The solution scheme described in the previous
section was used to obtain the equilibrium pair correlation function
$g(r,\rho)$ for the hard sphere fluid.  We compare the $g(r,\rho)$
obtained for different densities in Figs. \ref{g_r1}, \ref{g_r2} and
\ref{g_r3}. The same figures also compare the calculated values of
$g(r,\rho)$ with those obtained from molecular dynamics
simulations~\cite{Lab1}. In general, the ESPT prediction for
$g(r,\rho)$ is in very good agreement with molecular dynamics
calculations, indicating the success of the theory in capturing
fluid structure and its density dependence. However, for $\rho\geq
0.4$, the theory underpredicts $g(1,\rho)$, as will be discussed in
connection with equation of state predictions.

The computation time required for each successive calculation to
converge increases with density. In this work, we have been able to
continue the pair correlation calculations to densities that are in
the metastable region of the accepted hard sphere phase diagram. The
generally accepted freezing density for the hard sphere fluid is
$\rho_{freezing}=0.943$ ~\cite{Hoover1}. In Fig. \ref{0.96}, the
pair correlation function based on the ESPT for a density
$\rho=0.96$ is shown. The second peak appears to be almost
triangular in shape instead of the smooth peak observed at lower
densities. Note that the development of this near-triangular peak
shape is also evident in the molecular dynamics simulation results.
Calculations of the pair correlation function for $\rho>0.96$ using
the above discussed numerical scheme do not converge. Further
refinement of the numerical approach is under investigation.

\begin{figure}[!h]
\begin{center}
\includegraphics[width=8.8cm]{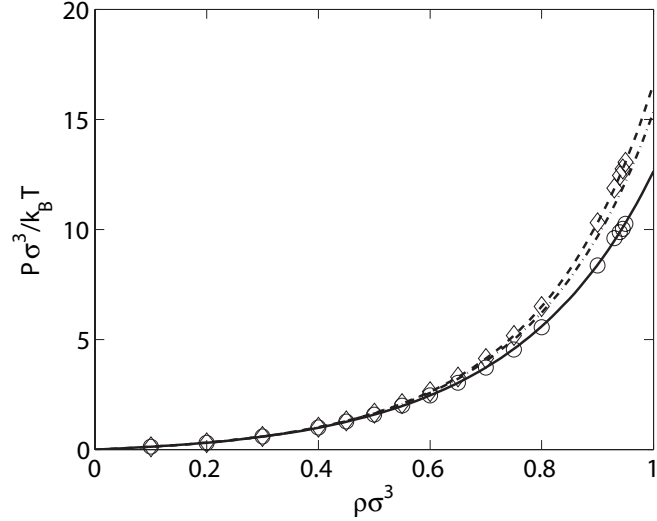}
\end{center}\caption{Equation of state of the hard sphere fluid. We compare predictions from ESPT using equation \ref{EOSeqnP} ($\circ$), the original SPT of Reiss et al.~\cite{Reiss1} ($\lozenge$), the Percus-Yevick~\cite{Per1} pressure (solid) and compressibility (dashed) calculations, and molecular dynamics simulation results~\cite{Kol1} (dot-dashed).}\label{EOS}
\end{figure}

The contact value $g(\sigma,\rho)$ can be used to calculate the
pressure of the fluid $P$ using the virial theorem,
\begin{equation}\label{EOSeqnP}
\beta P = \rho +(2/3)\pi\rho^{2}\sigma^3 g(\sigma,\rho)
\end{equation}
In Fig \ref{EOS}, the pressure is plotted against density for the
ESPT along with predictions from other equations of state (EOS's)
for the hard sphere fluid. The pressure prediction from the ESPT of
``I" corresponds closely to molecular simulation results~\cite{Kol1} at low densities. However at higher densities, there is
significant deviation between the two. In fact, we find that the
ESPT pressure is very close to the corresponding quantity computed
via the Percus-Yevick closure~\cite{Per1} (pressure equation).
This, in itself, is an interesting result in light of the fact
~\cite{Wert1} that the original SPT of Reiss et al.~\cite{Reiss1} produces the same EOS as the Percus-Yevick closure
(compressibility equation)~\cite{Per1}. Both versions of SPT
are unable to accurately predict the pressure at higher densities.
The original SPT over-estimates the pressure, while the ESPT
underestimates it.

\begin{figure}[!h]
\begin{center}
\includegraphics[width=8.8cm]{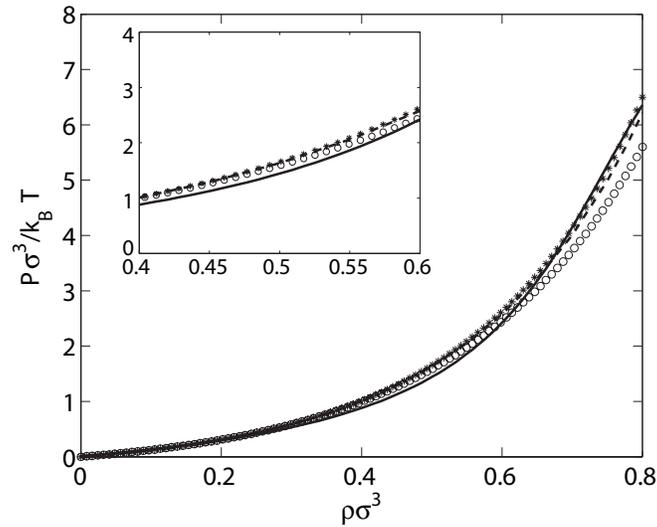}
\end{center}\caption{Equation of state of the hard sphere fluid We compare predictions from ESPT using the virial ($\circ$) and the compressibility (solid) routes, the original SPT of Reiss et al.~\cite{Reiss1} (*) and molecular dynamics simulation results~\cite{Kol1} (dashed). Inset shows details in the density range
$0.4\leq\rho\sigma^3\leq 0.6$.}\label{EOS2}
\end{figure}

An alternative calculation can be conducted using the isothermal
compressibility equation
\begin{equation}\label{EOSeqnC}
\beta\bigg(\frac{\partial \rho}{\partial
P}\bigg)_T=1+4\pi\rho\int[g(r,\rho)-1]r^2dr
\end{equation}
The derivative $(\partial P/\partial\rho)_T$ can then be integrated
numerically from the ideal gas limit, [$\beta P=\rho$; $g(r,\rho)=1$
for $r>1$, as $\rho\rightarrow 0$]. We use density steps of
$\delta\rho=.02$ in this calculation, ensuring that computed
dimensionless pressure values, $P\sigma^3/(k_BT)$, are accurate
within $10^{-2}$. A cut-off distance in $r$ is selected such that
contributions to the integral in equation \ref{EOSeqnC} are
considered only when $|g(r,\rho)-1|$ at any peak of $g(r,\rho)$ is
greater than $2\times 10^{-3}$. The pressure obtained by this route
is shown in Fig. \ref{EOS2}. In the figure, we compare the EOS
predictions from equations \ref{EOSeqnP} and \ref{EOSeqnC} with the
corresponding original SPT prediction~\cite{Reiss1}. At high
densities, the ESPT calculation via the compressibility equation
closely matches the original SPT, or equivalently~\cite{Wert1}, the
Percus-Yevick compressibility calculation. However, at low density,
there is significant difference between the two, with the ESPT
compressibility calculation yielding lower pressures. The same
figure with the accompanying inset also displays EOS values as
obtained from molecular simulation results. The EOS prediction from
ESPT through the compressibility route underpredicts the pressure at
lower densities ($\rho\leq 0.67$) and overpredicts this quantity at
higher densities ($\rho\geq 0.67$), when compared to simulation
results.

\begin{figure}[!h]
\begin{center}
\includegraphics[width=8.8cm]{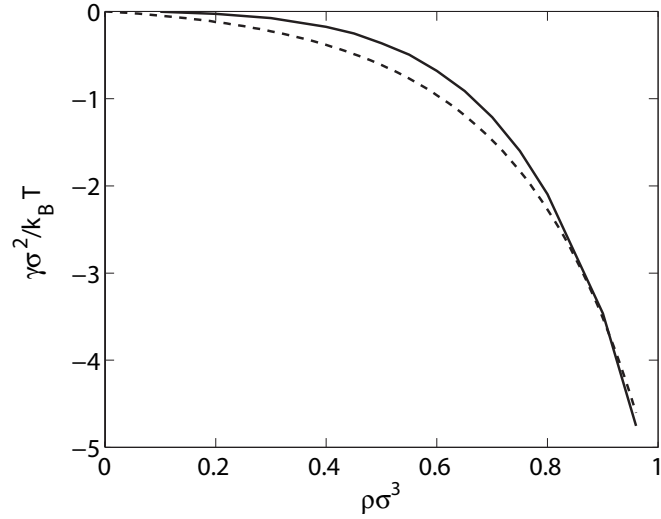}
\end{center}\caption{Surface tension of the hard sphere fluid. The solid line represents the prediction from ESPT; the dotted line, the original SPT of Reiss et al.~\cite{Reiss1}.}\label{G_rho}
\end{figure}

\begin{figure}[!h]
\begin{center}
\includegraphics[width=8.8cm]{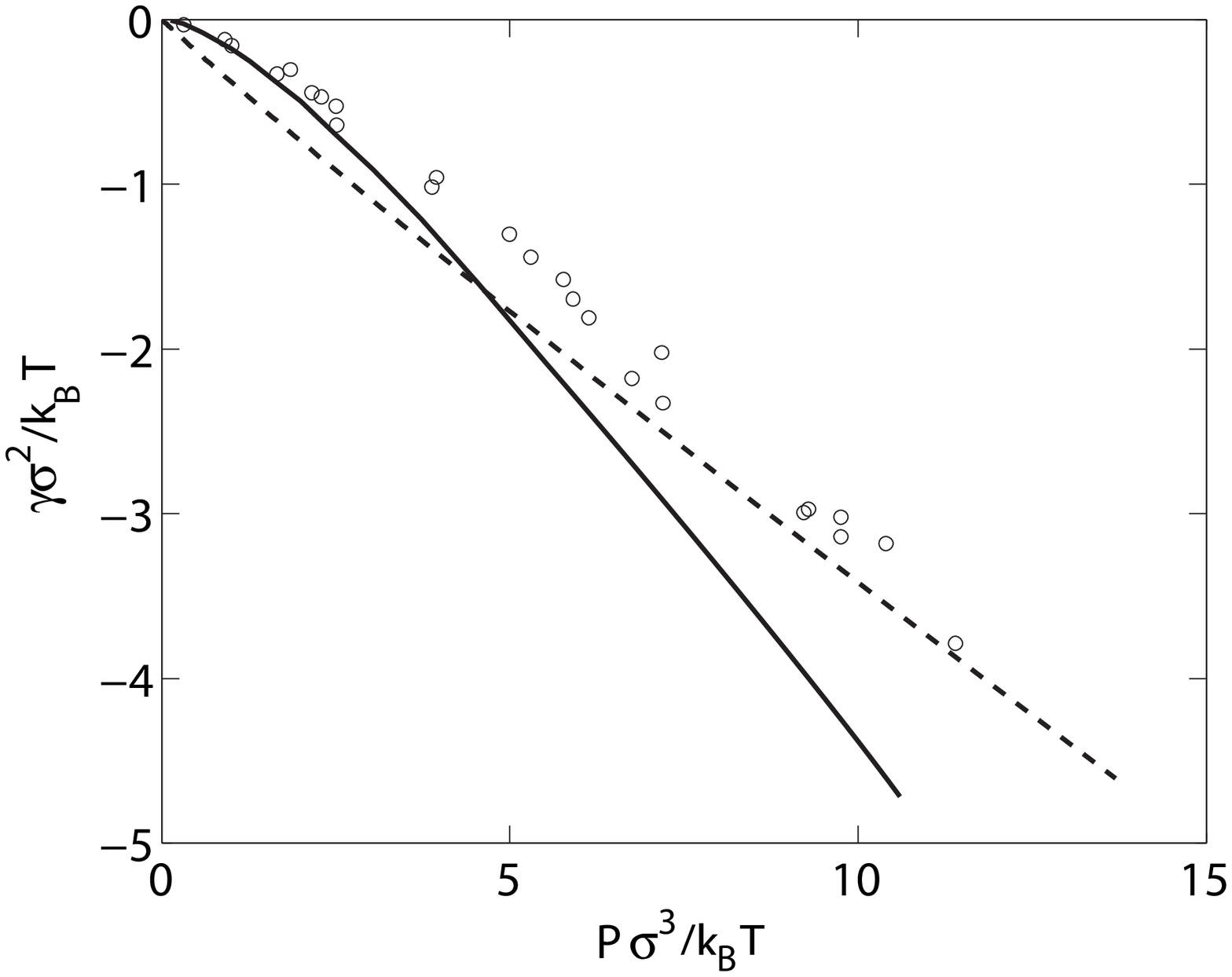}
\end{center}\caption{Surface tension of the hard sphere fluid. The solid line represents the prediction from ESPT using the virial route; the dotted line, the original SPT of Reiss et al.~\cite{Reiss1}. Open circles ($\circ$) are simulation results~\cite{Att1, Hen1, Hen2}.}\label{G_P}
\end{figure}

\begin{figure}[!h]
\begin{center}
\includegraphics[width=8.8cm]{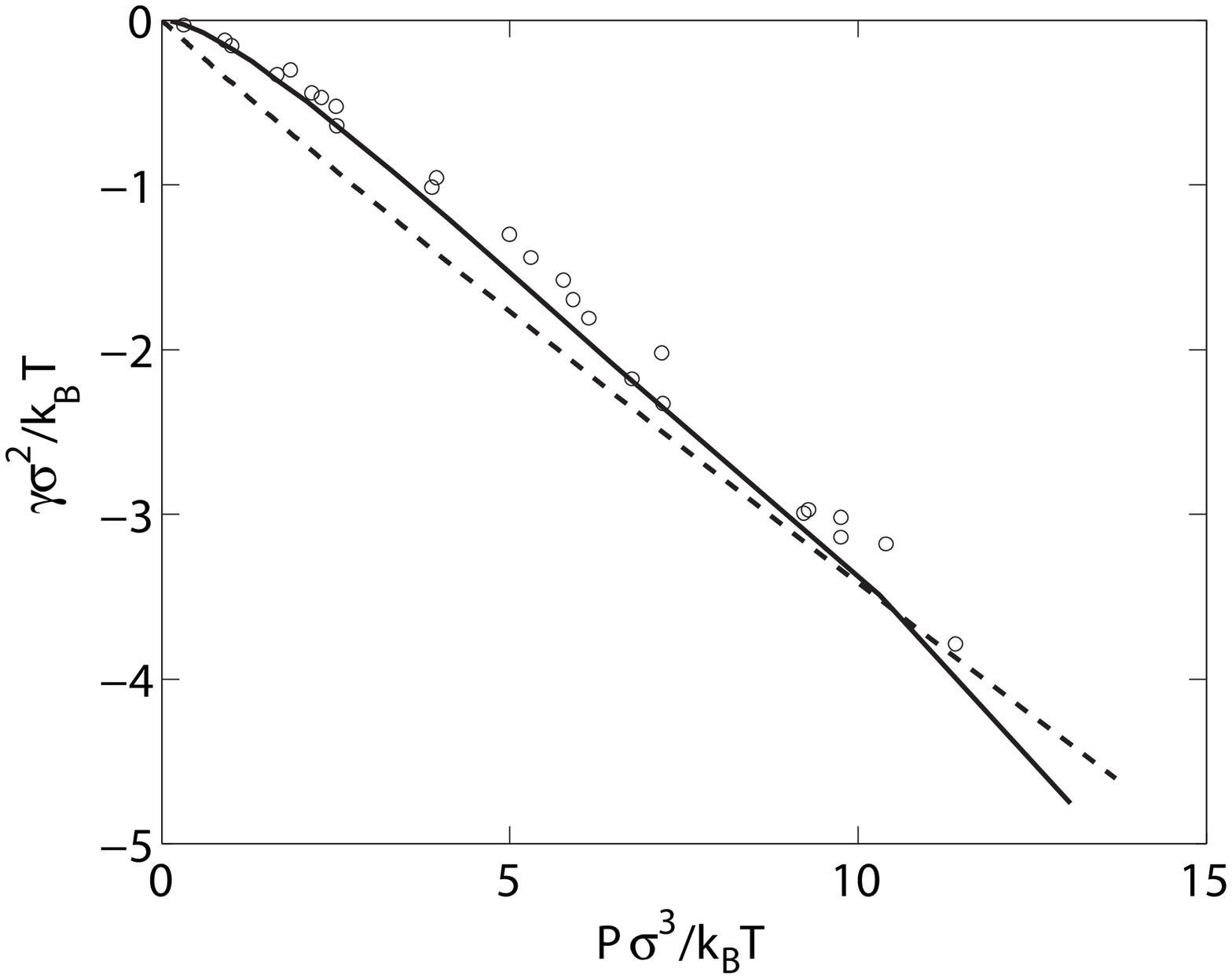}
\end{center}\caption{Surface tension of the hard sphere fluid. The solid line represents the prediction from ESPT using the compressibility route; the dotted line, the original SPT of Reiss et al.~\cite{Reiss1}. Open circles ($\circ$) are simulation results~\cite{Att1, Hen1, Hen2}.}\label{G_P2}
\end{figure}

The surface tension of the hard sphere fluid can be estimated from
the parameter $K$ in the asymptotic form of the cavity formation
work $\beta W(\lambda,\rho)$ in equation \ref{Wgt3}
\begin{equation}
\beta\gamma=\frac{K}{4\pi}
\end{equation}
where $\gamma$ is the surface tension. A comparison of $\gamma$ from
the ESPT with that of the original SPT is shown in Fig. \ref{G_rho}.
The original SPT has been shown to predict, with reasonable
accuracy, the surface tension of the hard sphere fluid, when
compared to simulation results~\cite{Sid2}. Both theories predict
negative surface tension for the the hard sphere fluid with $\gamma$
increasing in magnitude with increasing density. Similar values of
$\gamma$ are predicted by both theories at low and high densities.
However, for intermediate densities, the ESPT predicts a smaller
magnitude for the surface tension than the original SPT. Comparison
with simulation results~\cite{Att1, Hen1, Hen2} is shown in Fig.
\ref{G_P}. Here, the pressure prediction for the different theories
and simulations is plotted on the abscissa. We observe that the ESPT
is able to provide a more accurate prediction of the surface tension
at low density and pressure. However as the density increases, the
original SPT is able to more accurately capture the dependence of
surface tension on pressure. Equation \ref{EOSeqnC} provides an
alternate route for the calculation of the EOS using the ESPT. As
shown in Fig. \ref{G_P2}, use of the compressibility route enables
the ESPT to provide an improved representation of simulation results
for the variation of surface tension with pressure than the original
SPT.

\begin{figure}[!h]
\begin{center}
\includegraphics[width=8.8cm]{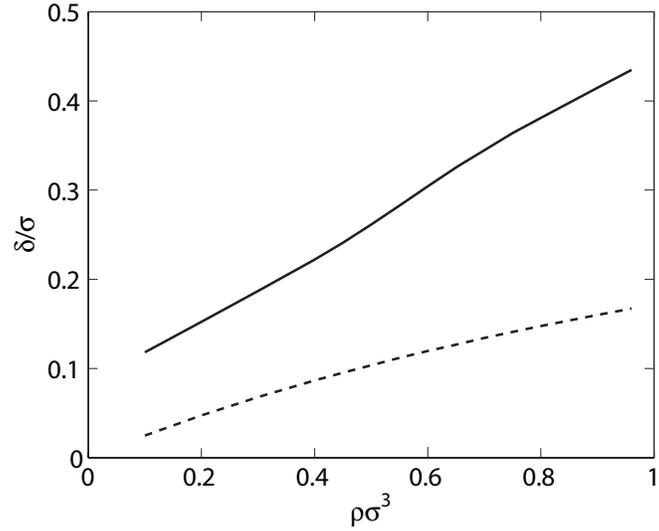}
\end{center}\caption{Tolman length of the hard sphere fluid as predicted by ESPT (solid) and the orginal SPT of Reiss et al.~\cite{Reiss1} (dashed).}\label{Delta}
\end{figure}

The Tolman length $\delta$ is a measure of the dependence of surface
tension on curvature. It can be estimated from parameter $L$ in the
asymptotic equation \ref{Wgt3} from the following relation
\begin{equation}
\delta=-\frac{Lk_B T}{8\pi\gamma}
\end{equation}
For a hard sphere fluid $\delta>0$. The predicted variation of
$\delta$ with density is shown in Fig. \ref{Delta}. The Tolman
length increases monotonically upon compression. Comparing these
values with those from the original SPT, we find that ESPT predicts
a more pronounced dependence of the surface tension on curvature.

\section{Conclusions}
A new scaled particle theory of hard sphere pairs~\cite{Stillinger1} has been used to calculated the pair correlation function
$g(r,\rho)$ of the hard sphere fluid over a wide range of densities,
covering the generally accepted stable and slightly metastable fluid
states. Comparison with results from molecular dynamics
simulations~\cite{Lab1} shows that the extended scaled particle
theory (ESPT) is able to accurately describe the fluid's structure,
but it underestimates the $g(r,\rho)$ contact value when $\rho>0.4$.
The theory is able to accurately reproduce the shape of the peak
corresponding to second neighbor shell at $r\cong2$ which becomes
sharper as the density of the hard sphere fluid increases, in
agreement with simulation results.

The ESPT has been used to compute the equation of state of the hard
sphere fluid by using the virial theorem. The calculated EOS values
correspond closely at lower densities with those obtained from
molecular simulations~\cite{Kol1}. However, there are significant
differences between the two at higher densities. We find that ESPT
consistently underpredicts EOS values from molecular simulation.
This can be contrasted against the overprediction of the EOS by the
original SPT. Comparing the predictions of ESPT with those from
other theories, we find that the calculated EOS values closely match
those computed using the Percus-Yevick approximation (pressure
equation). We have also calculated the hard sphere equation of state
using ESPT with the compressibility equation. The resulting
prediction is close to the equation of state obtained using the
Percus-Yevick closure and the compressibility equation, especially
at high densities.

The surface tension and Tolman length have also been calculated. As
expected, ESPT predicts a negative surface tension for the hard
sphere fluid. ESPT is able to more closely capture the pressure
dependence of the surface tension at moderate pressures than the
original SPT~\cite{Reiss1}. However, at higher pressures, the
original SPT provides better prediction. As discussed above,
employing the compressibility route improves the EOS prediction of
ESPT, leading to a closer correspondence with $\gamma$ vs. $P$
surface tension predictions from molecular simulation. ESPT predicts
a larger Tolman length than the original SPT, or in other words, the
new theory predicts a more pronounced effect of curvature on surface
tension than the original theory.

Having studied a small portion of the metastable branch of the
equation of state, future research directions might include an
examination of the high density range of the phase diagram in an
attempt to calculate the pair correlation function of both the hard
sphere solid and the metastable fluid. Using solid and fluid forms
of $g(r,\rho)$ as initial inputs to the calculation is a possible
strategy that will be investigated as a means of testing the present
theory's ability to predict a freezing transition. Further numerical
work is needed to ensure convergence along the high-density
metastable branch.

Another avenue of investigation is to extend the current hard sphere
theory to study solvation effects. The new theory would include the
effect of orientation of and attraction between interacting
particles. There have been other attempts at developing such
theories in the past~\cite{Ash1, Stillinger2}.
These have been successful in highlighting important phenomena
associated with hydrophobic hydration or solvation.

Finally, we point out that additional consistency conditions can be
imposed in the iteration scheme described in Section III. The extent
to which such additional constraints improve the accuracy of ESPT
predictions of thermodynamic quantities such as the pressure and the
surface tension deserves further investigation.

\section*{ACKNOWLEDGEMENTS}
P.G.D. gratefully acknowledges financial support by the U.S.
Department of Energy, Division of Chemical Sciences, Geosciences,
and Biosciences, Office of Basic Energy Sciences, grant no.
DE-FG02-87ER13714. The authors gratefully acknowledge Yannis
Kevrekedis and Liang Qiao for their help with the numerical solution
of the non-linear integral equation.

\newpage
\bibliography{Man}

\end{document}